\documentclass[pra,twocolumn,nofootinbib,floatfix,10pt]{revtex4-2}

\usepackage{amsmath}
\usepackage{amssymb}
\usepackage{wasysym}
\usepackage{graphicx}
\usepackage{color,soul}
\usepackage{physics}
\usepackage{siunitx}
\usepackage{dsfont}
\usepackage{float}
\usepackage{xcolor}
\usepackage[english]{babel}
\usepackage{blindtext}
\usepackage[english,nomargin,inline,marginclue,draft]{fixme}
\pdfpageheight\paperheight
\pdfpagewidth\paperwidth

\usepackage[colorlinks,linkcolor=blue,anchorcolor=blue,citecolor=blue,urlcolor=blue]{hyperref}

\fxusetheme{colorsig}
\FXRegisterAuthor{cg}{acg}{CG}  
\FXRegisterAuthor{th}{ath}{\color{blue}TH}  
\FXRegisterAuthor{ib}{aib}{\color{red}IB} 
\FXRegisterAuthor{sh}{ash}{\color{cyan}SH} 
\FXRegisterAuthor{db}{adb}{\color{green}DB} 
\FXRegisterAuthor{ps}{aps}{PS}
\makeatletter
\renewcommand*\FXLayoutInline[3]{%
  {\@fxuseface{inline}\ignorespaces{\color{fx#1}[#3: #2]}}}
\makeatother

\long\def\symbolfootnote[#1]#2{\begingroup%
\def\thefootnote{\fnsymbol{footnote}}\footnotetext[#1]{#2}\endgroup}

\def\nobreakbefore{%
  \relax\ifvmode\else
    \ifhmode
      \ifdim\lastskip > 0pt\relax
        \unskip\nobreakspace
      \else 
        \nobreakspace
      \fi
    \fi
  \fi
}
\let\oldcite\cite
\renewcommand\cite{\nobreakbefore\oldcite}

\begin{document}
\title{Exceptional point and hysteresis trajectories in cold Rydberg atomic gases}

\author{Jun Zhang$^{1,2,\textcolor{blue}{\dagger}}$}
\author{En-Ze Li$^{1,2,\textcolor{blue}{\dagger}}$}
\author{Ya-Jun Wang$^{1,2,\textcolor{blue}{\dagger}}$}
\author{Bang Liu$^{1,2}$}
\author{Li-Hua Zhang$^{1,2}$}
\author{Zheng-Yuan Zhang$^{1,2}$}
\author{Shi-Yao Shao$^{1,2}$}
\author{Qing Li$^{1,2}$}
\author{Han-Chao Chen$^{1,2}$}
\author{Yu Ma$^{1,2}$}
\author{Tian-Yu Han$^{1,2}$}
\author{Qi-Feng Wang$^{1,2}$}
\author{Jia-Dou Nan$^{1,2}$}
\author{Yi-Ming Ying$^{1,2}$}
\author{Dong-Yang Zhu$^{1,2}$}
\author{Bao-Sen Shi$^{1,2}$}
\author{Dong-Sheng Ding$^{1,2,\textcolor{blue}{\star}}$}

\affiliation{$^1$Key Laboratory of Quantum Information, University of Science and Technology of China, Hefei, Anhui 230026, China.}
\affiliation{$^2$Synergetic Innovation Center of Quantum Information and Quantum Physics, University of Science and Technology of China, Hefei, Anhui 230026, China.}

\date{\today}

\symbolfootnote[2]{J.Z, E.Z.L, and Y.J.W contribute equally to this work.}
\symbolfootnote[1]{dds@ustc.edu.cn}

\maketitle

\textbf{The interplay between strong long-range interactions and the coherent driving contribute to the formation of complex patterns, symmetry, and novel phases of matter in many-body systems. However, long-range interactions may induce an additional dissipation channel, resulting in non-Hermitian many-body dynamics and the emergence of exceptional points in spectrum. Here, we report experimental observation of interaction-induced exceptional points in cold Rydberg atomic gases, revealing the breaking of charge-conjugation parity symmetry. By measuring the transmission spectrum under increasing and decreasing probe intensity, the interaction-induced hysteresis trajectories are observed, which give rise to non-Hermitian dynamics. We record the area enclosed by hysteresis loops and investigate the dynamics of hysteresis loops. The reported exceptional points and hysteresis trajectories in cold Rydberg atomic gases provide valuable insights into the underlying non-Hermitian physics in many-body systems, allowing us to study the interplay between long-range interactions and non-Hermiticity.}

Exceptional points (EPs) are special points in the parameter space of a non-Hermitian system where two or more eigenstates and their corresponding eigenvalues coalesce. At these points, compared to the Hermitian Hamiltonian, the eigenvalues of the underlying system's Hamiltonian have complex values \cite{miri2019exceptional,mandal2021symmetry,wang2022non,ding2022non,yoshida2023fate,ding2022non}. Recent studies have shown that open systems undergo phase transitions at EPs, leading to a variety of interesting physical phenomena, including chirality \cite{wang2020electromagnetically,shu2024chiral}, unidirectional transmission or reflection\cite{huang2017unidirectional,du2022light}, topological phase transition\cite{ding2024electrically,gong2024topological}, parity-time symmetry breaking \cite{el2018non,li2023exceptional} and charge-conjugation parity (CP) symmetry breaking \cite{delplace2021symmetry}, and supernormal sensitivity to perturbations \cite{rosa2021exceptional,yang2023spectral}. These properties have become the focus of research on non-Hermitian systems associated with EPs, which opened the door to a series of experimental studies in optics\cite{peng2014loss,hodaei2014parity,wang2023experimental}, electronics\cite{chitsazi2017experimental,assawaworrarit2017robust}, and enhanced sensing \cite{chen2017exceptional,lau2018fundamental,budich2020non,li2023synergetic,liang2023observation} (where the sensor sensitivity is enhanced by energy bifurcation near the EPs). The combination between non-Hermiticity and many-body interaction enables the emergence of non-trivial effects \cite{hamazaki2019non,matsumoto2020continuous,kawabata2022many}, thus providing a platform to study the emergent phases beyond few-body scenarios. 

Due to the strong dipole interaction between Rydberg atoms \cite{saffman2010quantum,adams2019rydberg,browaeys2020many}, they have become a versatile tool for studying many-body physics. This long-range interaction induces non-linearity and gives rise to a unique dissipation channel, enabling investigations into non-equilibrium phase transitions \cite{lee2012collective,carr2013nonequilibrium,schempp2014full,marcuzzi2014universal,lesanovsky2014out, urvoy2015strongly, ding2022enhanced}, self-organization \cite{Signatures2020Helmrich,ding2019Phase,Wintermantel2020Cellular, klocke2021hydrodynamic}, ergodicity breaking and time crystals \cite{gambetta2019, Wadenpfuhl2023Synchronization, ding2023ergodicity, wu2023,liu2024higher,liu2024bifurcation,liu2024microwave}. The dissipation induced by interactions can cause energy to be exchanged between a many-body system and its external environment, thereby influencing new dynamics and symmetric properties of the system, i.e. non-Hermitian many-body physics. Thus, studying the relationship between many-body interactions and non-Hermiticity can provide a new framework to investigate non-Hermitian dynamics in many-body scenarios \cite{pan2020non}. Studying these in Rydberg atom systems has advantages in precise control of interactions, offering valuable insights in finding new emergent phases in symmetry breaking \cite{delplace2021symmetry}, which is the starting point of this work.

Here, we propose a new paradigm for studying non-Hermitian physics by considering interaction-induced dissipation in cold Rydberg atomic gases. When the interaction between Rydberg atoms is weak and there is a normal electromagnetically induced transparency (EIT) spectrum, whilst the peak of EIT splits into two when in strong interaction, indicating that the system crosses the third-order EPs. In this scenario, the interaction is a dominant resource for system to produce non-Hermitian features, leading to rich hysteresis trajectories by varying probe intensities. The area of hysteresis loops reveals energy loss due to non-Hermiticity, and the dynamics can be tuned in various time scales.

\begin{figure*}
\centering
\includegraphics[width=1\linewidth]{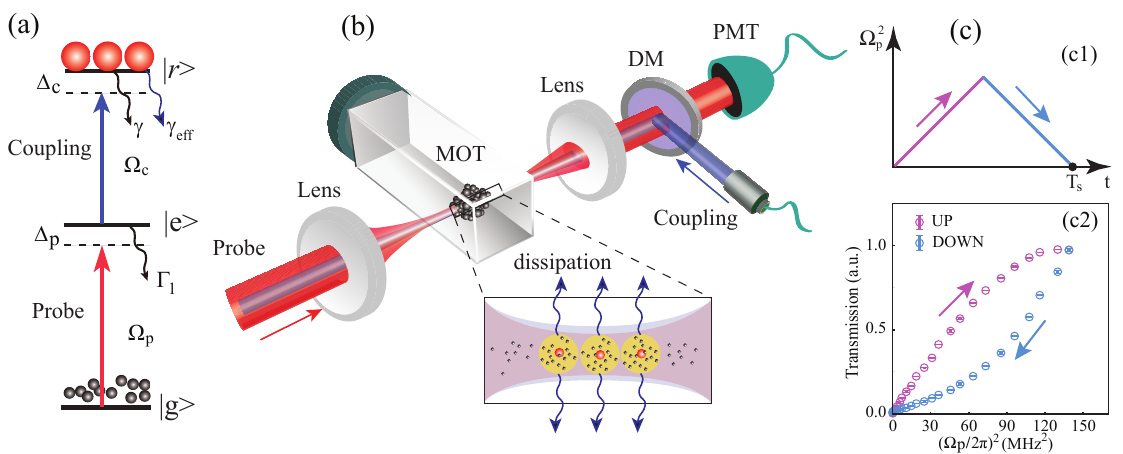}
\caption{\textbf{Schematic of many-body interaction induced EPs and hysteresis loops in Rydberg atoms.} 
(a) Rydberg atomic energy level diagrams. Probe $\Omega_{p}$ and coupling $\Omega_{c}$ fields with detunings $\Delta_{p}$ and $\Delta_{c}$. $\Gamma_{1}$ and $\gamma$ are decay rates of states $\ket{e}$ and $\ket{r}$, and $\gamma_{\text{eff}}$ is the decay rate caused by Rydberg many-body interactions. 
(b) Schematic diagram of the experimental setup. The probe beam is incident opposite to the coupling beam through the lens and focused in a magneto-optic trap (MOT) trapping $^{85}Rb$ atoms. (c) Measured transmission by positively (Up, pink) and negatively (Down, blue) scanning $(\Omega_{p}/2\pi)^{2}$, the trajectories connected by data points exhibit hysteresis loop.}
\label{Fig1}
\end{figure*}

\subsection*{Physical Model}
Our model is based on a three-level Rydberg atomic system, as depicted in Fig.~\ref{Fig1}(a). There are three atomic state manifolds of ground state $\ket{g}$, metastable state $\ket{e}$, and Rydberg state $\ket{r}$. The probe field with Rabi frequency (detuning) $\Omega_{p}$ ($\Delta_{p}$) drives the transition $\ket{g}\leftrightarrow\ket{e}$, and the coupling field with Rabi frequency (detuning) $\Omega_{c}$ ($\Delta_{c}$) drives the transition $\ket{e}\leftrightarrow\ket{r}$. The spontaneous decay rates of the states $\ket{e}$ and $\ket{r}$ are $\Gamma_1$ and $\gamma$, respectively. A pair of atoms $i$ and $j$ at positions $r_i$ and $r_j$ excited to the Rydberg states $\ket{r}$ interact with each other via a van der Waals (vdW) potential $V_{\text{vdW}}\propto{C_6}/{R^6}$, where $C_6$ is the coefficient. 

By the EIT theory of the ensemble of cold atoms, long-range interactions between atoms limit the medium to behave as a collection of superatoms (Rydberg polaritons), each containing a blockade volume that can hold at most one Rydberg excitation as shown in Fig.\ref{Fig1}(b). The experimental setup is depicted by Fig.\ref{Fig1}(b), and the scan of $\Omega_{p}^2$ probes the dynamics of system response, as given in Fig.\ref{Fig1} (c). The Rydberg polaritons display a dephasing feature when considering the interactions with each other, where the non-uniform distribution of the Rydberg atoms inside the polariton causes the position-dependent phase shifts \cite{busche2017contactless,bariani2012dephasing}. 

The interaction between polaritons accelerates the decay of Rydberg atoms and causes a broadening of the Rydberg energy levels. This creates an additional dissipation channel for the Rydberg state \(\ket{r}\) due to its surrounding environment; further analysis can be found in the Methods section. In the rotating frame, the Hamiltonian of our system takes the form
\begin{equation}
H_1=
\left(
\begin{array}{ccc}
0           & \Omega_{p}/2    &       0     \\
\Omega_{p}/2  & \Delta_{p}    &   \Omega_{c}/2\\
0           & \Omega_{c}/2    &   \Delta_c + \Delta_p - i\gamma_{\rm{eff}}/2
\end{array}
\right),
\label{eq1}
\end{equation}
where $\gamma_{\rm{eff}}$ donates the effective non-Hermitian term. Obviously, when $\Delta_p=\Delta_c=0$, this system is protected by CP-symmetry, and the non-Hermitian Hamiltonian satisfies ${U_{CP}}{H_1}U_{CP}^{-1}=-{{H_1}^*}$ with ${U_{CP}}={\rm{diag}}(1,-1,1)$ \cite{delplace2021symmetry}. The eigenvalues of $H_{\text{1}}$ are calculated as $E$ = $E_1$, $E_2$, $E_3$, see more details in the Methods section. 

\begin{figure*}
\centering
\includegraphics[width=1\linewidth]{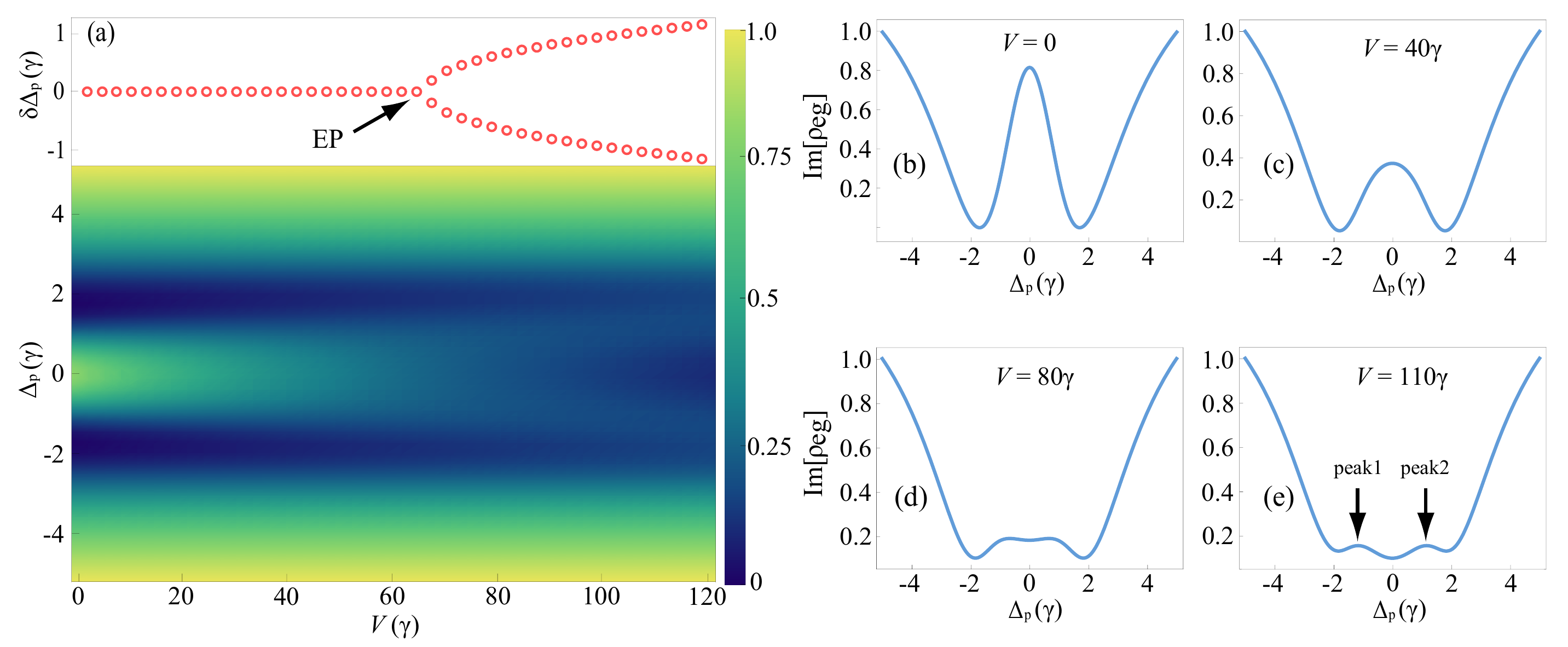}
\caption{\textbf{Theoretical phase diagram.}  (a) Theoretical spectrum Im$[\rho_{eg}]$ versus the interaction strength $V$ and detuning $\Delta_p$. As the interaction strength gradually increases, Im$[\rho_{eg}]$ decreases, as shown in (b) and (c). With further increase in interaction strength, the system transits an EP (marked by the black arrow in the up panel of (a)) and two distinct transmission peaks emerge [marked by the two black arrows] in the spectrum, as depicted in (d) and (e). The frequency difference $\delta$ between these emergent two peaks is indicated by the red circles in the up panel in (a). A value of 0 means that the system does not exhibit two distinguishable peaks. In these simulations, we set $\Gamma_1 = 6\gamma$, $\Delta_c$=0, $\Omega_c = 3 \gamma$, and $\Omega_p=\gamma$. }
\label{Fig2.2}
\end{figure*}

In this case, we can observe the Riemann surface showing real and imaginary parts of the eigenvalues of Hamiltonian $H_{1}$ [Fig.~\ref{FigS1}(a-b) in Methods section]. The non-zero $\gamma_{\rm{eff}}$ induces imaginary parts of eigenvalues and generates the non-Hermitian features. By increasing $\gamma_{\rm{eff}}$, the coalesce of three eigenvalues emerges and results in complex singularities in energy spectrum [see more information in Figs.~\ref{FigS1}(c-f)], we called these points as third-order EPs \cite{mandal2021symmetry,delplace2021symmetry}.

At the third-order EPs, the real and imaginary parts of the eigenvalues behave non-trivial, which are different from the case at the second-order EPs in PT-symmetry two-level system where the real and imaginary parts of the eigenvalues will degenerate simultaneously \cite{konotop2016nonlinear,el2018non,mandal2021symmetry}. In the scenario of third-order EPs, a triple degeneracy of the eigenenergy exists in real space ($\text{Re}[E_{i}]=0$, $i\in\{1,2,3\}$), the corresponding imaginary parts separate into three values [see the gray regimes in Figs.~\ref{FigS1}(c-f) in Methods section, in which $E_{1}\neq E_{2}\neq E_{3}$ and $ E_{i}\in i\mathbb{R}$, indicating CP-symmetry breaking \cite{delplace2021symmetry}. When the eigenvalues of real space are completely non-degenerate, the eigenvalues of imaginary space separate into two values where two of the three are degenerate [here, $E_{1}=-E_{2}^{*}$ and $E_{3}\in i\mathbb{R}$], see the blue regions in Figs.~\ref{FigS1}(c-f) in Methods section.

We calculate the solution of master equation $\dot{\rho}=-i[\hat{H_1},\rho]+\mathcal{L}[\rho]$ at the steady-state condition $\dot{\rho}=0$. By the treatment of mean-field field, we map the spectrum of Im$[\rho_{eg}]$ versus the interaction strength $V$ and detuning $\Delta_p$, as shown by the down panel in Fig.~\ref{Fig2.2}(a). With increase of $V$, for example, from $V=0$ to $V=110\gamma$, the peak of Im$[\rho_{eg}]$ decreases and the neighbor two peaks emerge, as shown in Figs.~\ref{Fig2.2}(b-e) and the up panel in Fig.~\ref{Fig2.2}(a). In this process, the system crosses the EPs, and the peak of Im$[\rho_{eg}]$ splits into two,  indicating that the degeneracy of one imaginary eigenvalue of system has been de-degenerated.

\subsection*{Non-Hermitian spectrum and exceptional point}
\begin{figure*}
\centering
\includegraphics[width=1\linewidth]{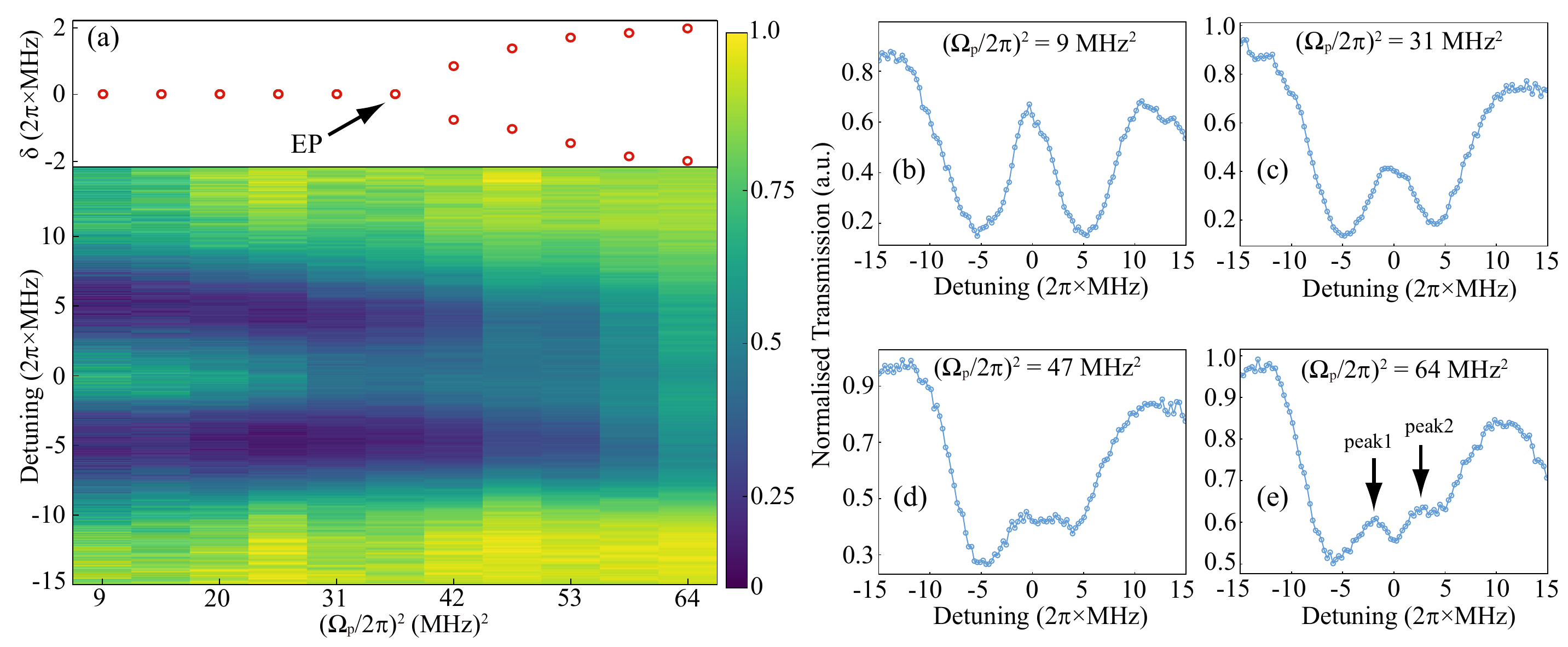}
\caption{\textbf{Measured phase diagram.} (a) Measured EIT transmission spectrum versus the probe intensity $\Omega_{p}^{2}$. The EIT spectrum shows a visible transmission peak when the probe intensity is low, which means that the interaction between Rydberg atoms are so weak that the system maintains Hermiticity, as shown in (b). As the probe intensity grows, non-Hermiticity of the Rydberg atoms system begins to emerge, which is manifested in the transmission spectrum as a weakening of the transmission intensity, as shown in (c). And as the probe intensity continues to increase, the strong interaction between the Rydberg atoms brings out the eigenenergy bifurcation, and two transmission peaks appear in the EIT spectrum, as shown in (d).}
\label{Fig2}
\end{figure*}

The presence of strong interactions between Rydberg atoms leads to additional dissipation, which provides a platform to study non-Hermitian spectrum. In the experiment, we excite the ground state of a cold ensemble of $^{85}$Rb atoms to Rydberg state $\ket{47D_{5/2}, F = 5, m_{F} = -5}$ by the EIT method, see more detailed information in Methods section. We measure the spectra by scanning probe detuning  $\Delta_{p}$ from $\Delta_{p}$ = -2$\pi\times$15 MHz to $\Delta_{p}$ = 2$\pi\times$15 MHz under different probe intensities $(\Omega_{p}/2\pi)^{2}$. By this way, we can map a phase diagram of system that responds to parameters of probe detuning $\Delta_{p}$ and intensity $(\Omega_{p}/2\pi)^{2}$.  From $(\Omega_{p}/2\pi)^{2}$ = 9 MHz$^{2}$ to $(\Omega_{p}/2\pi)^{2}$ = 64 MHz$^{2}$, the dynamics of response are obtained. When the probe intensity $(\Omega_{p}/2\pi)^{2}$ is small [for example, $(\Omega_{p}/2\pi)^{2}\leq 14.5$ MHz$^{2}$], the EIT spectrum is normal as the interaction between Rydberg atoms is weak. In this process, the system has three eigenvalues as we can see only one peak and two dips in Fig.\ref{Fig2}(b) [the peak and dips result from one zero energy and two symmetric eigenenergy as described in \cite{fleischhauer2005electromagnetically}], and this corresponds to the regime of Hermiticity approximately [as the interaction is ignored].

When the probe intensity $(\Omega_{p}/2\pi)^{2}$ is large [for example, $(\Omega_{p}/2\pi)^{2}> 14.5$ MHz$^{2}$], non-Hermiticity of system begins to emerge and the peak intensity of the EIT spectrum becomes weak, as shown in Fig.\ref{Fig2}(c). With a further increase of $(\Omega_{p}/2\pi)^{2}$, the peak of EIT spectrum splits into two peaks and the system undergoes the third-order EPs [see the results in Figs.\ref{Fig2}(d) and (e)]. At EPs (($\Omega_{p}/2\pi)^{2}$=36.5 MHz$^{2}$), a sudden change in the physical parameters breaks symmetry of system, leading to a bifurcation where the real and imaginary parts of system eigenvalue coalesce and split respectively. In this scenario, the presence of spectrum splitting post-EPs signatures the breaking of CP-symmetry \cite{delplace2021symmetry}. The peaks in Fig.\ref{Fig2}(e) are asymmetric due to the small shift on Rydberg energy level. 

\subsection*{Hysteresis trajectories.}
\begin{figure*}
\centering
\includegraphics[width=1\linewidth]{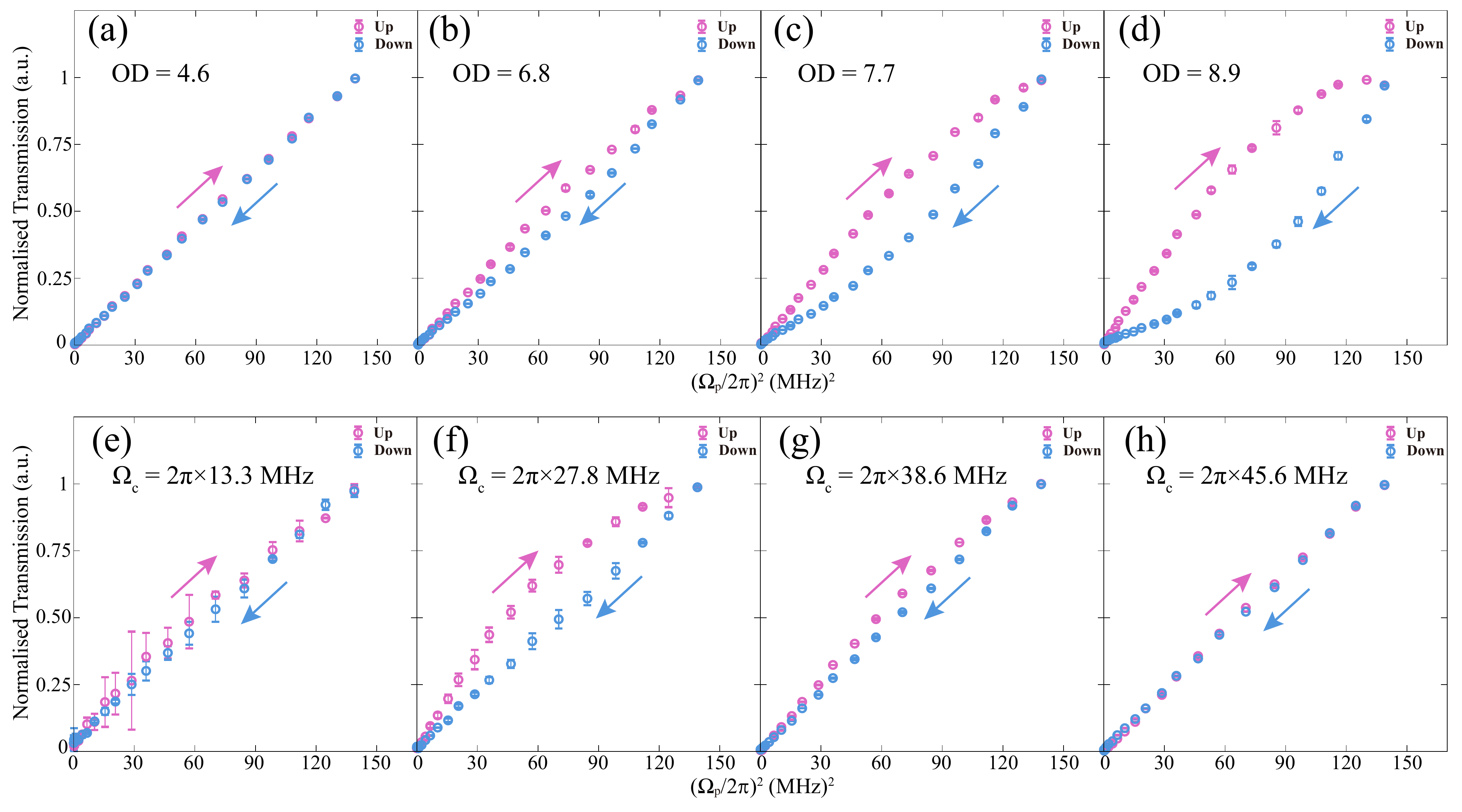}
\caption{\textbf{Measured hysteresis trajectories versus OD and $\Omega_{c} $.} (a)-(d) Measured transmission with scanning $\Omega_p$ in positive (pink, up) and negative (blue, down) directions under optical densities of OD = 4.6 (a), OD = 6.8 (b), OD = 7.7 (c) and OD = 8.9 (d). The pink circle data represents the transmission when increasing $(\Omega_{p} / 2\pi)^{2}$ from 0 MHz$^{2}$ to 140.6 MHz$^{2}$, and the blue circle data shows the transmission by scanning $(\Omega_{p} / 2\pi)^{2}$ from 140.6 MHz$^{2}$ to 0 MHz$^{2}$ with the same sweep rate.
(e)-(f) Measured transmission versus the coupling field's Rabi frequency of $\Omega_{c}$ = 2$\pi \times$13.3 MHz (e), $\Omega _{c}$ = 2$\pi \times$27.8 MHz (f), $\Omega _{c}$ = 2$\pi \times$38.6 MHz (g) and $\Omega _{c}$ = 2$\pi \times$45.6 MHz (h).}
\label{Fig3}
\end{figure*}

The dependence on the transmission on the probe intensity allows us to observe hysteresis trajectories, which reflects the interplay between Rydberg atoms' response and theirs interaction. By scanning $(\Omega_{p} / 2\pi)^{2}$ from 0 MHz$^{2}$ to 140.6 MHz$^{2}$ and vice versus, we can obtain a series of closed circles under different optical densities (ODs), as shown in in Figs.\ref{Fig3}(a)-(d). In the regime with small atoms numbers (which corresponds to large atomic distance $R$), we can find that the invariance of the linearity of transmission to probe intensity. This implies the interaction-induced dissipation does not dominate by comparing the inherent decay rate $\gamma$ of Rydberg state, corresponding to the physical process under the regime of Hermiticity, as given by the case of OD = 4.6 in Fig.\ref{Fig3}(a). When we increase OD, for example, from OD = 6.8 to OD = 8.9, the transmission for positive and negative scans undergoes behavior with different scaling. The physics behind this phenomenon is non-Hermiticity: at larger OD, the Rydberg atomic interaction-induced dissipation cannot be ignored and the response is no longer linear to probe intensity and results in the emergence of hysteresis loop, see the area between pink and blue data given in Figs.\ref{Fig3}(b)-(d).

We also model the trajectories of system using the Lindblad master equation $\dot{\rho}=-i[\hat{H_1},\rho]+\mathcal{L}[\rho]$, where the operator $\mathcal{L}$ describes the spontaneous emission rate of system. The theoretical results given in Figs.\ref{FigS2}(a-c) predict the hysteresis loops of Im[$\rho_{eg}$], see more details in Method sections. The phenomenon of hysteresis loop is counter-intuitive because the transmission does not overlap under the same probe intensity, in which the current state of system not only depends on its current inputs but also on its past states and inputs. The increase of atoms number through probe intensity forms structured Rydberg clusters by near neighbor interaction, the induced dissipation makes the clusters' behaviour different from the case of few atoms. If the probe intensity is then reduced conversely, the atoms retains some level of dissipation. To completely release the dissipation of atoms, only small reduction of probe intensity is required in the reversal scanning, see more detailed information in Methods section. This asymmetry in the response of the Rydberg atoms to increasing and decreasing probe intensity is a clear manifestation of hysteresis. 

The direction of hysteresis trajectories is reversal to magnetic hysteresis in ferromagnetic material \cite{jiles1986theory}, but same with the normal elastic hysteresis of rubber \cite{ogden2004elasticity}. The different directions of trajectories result from the distinct physical mechanism behind these hysteresis. In our experiment, different ODs (which relate to different numbers of atoms and interaction strengths) can alter how the system responds to changes in probe intensity, and influences the size (area) and shape of the hysteresis loop observed experimentally, as illustrated in Figs.\ref{Fig3}(a-d). In terms of energy, the area enclosed by the hysteresis loop quantifies the energy loss during scanning $(\Omega_{p} / 2\pi)^{2}$. 

The hysteresis loop also depends on the coupling Rabi frequency $\Omega_{c}$, as given in Figs.\ref{Fig3}(e)-(h). We can find that the hysteresis loop only appears within a range of $\Omega_{c}$. When we use a relative small Rabi frequency of $\Omega_{c}$ = 2$\pi \times$13.3 MHz, the corresponding excited Rydberg atoms number cannot provide enough interaction strength between atoms, thus resulting in normal trajectories shown in Fig.\ref{Fig3}(e). When we set $\Omega_{c}$ = 2$\pi \times$27.8 MHz, the interaction-induced dissipation affects the scaling of transmission to probe intensity, then generates a hysteresis loop given in Fig.\ref{Fig3}(f). If the system is driven under a large $\Omega_{c}$ (for example, $\Omega_{c}$ = 2$\pi \times$38.6 MHz and  $\Omega_{c}$ = 2$\pi \times$45.6 MHz) that the atoms have no time to respond, then the hysteresis loop disappears.

\subsection*{Hysteresis loops dynamics}
\begin{figure*}
\centering
\includegraphics[width=1\linewidth]{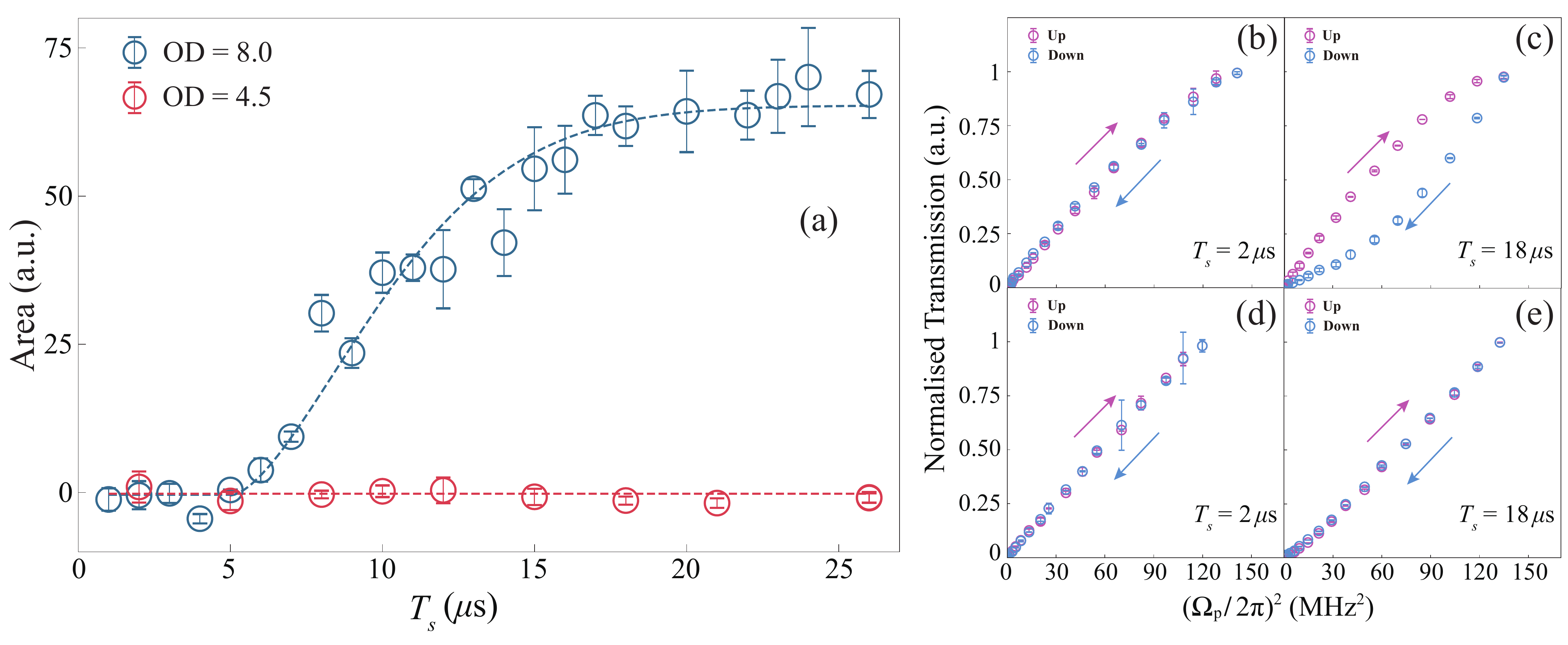}
\caption{\textbf{Hysteresis dynamics.} (a) The measured area of hysteresis loops versus the scan time  $T_{s}$ under OD = 8 (blue circles) and OD = 4.5 (red circles). The blue data is fit by the function of $y = ae^{-b(x-d)^{\alpha}}$+$c$ ($a$ = -64.9, $b$ = 0.053, $c$ = 65.3, $d$ = 5, $\alpha$ = 1.5), and the red data is fit by the function of $y = 0$. (b) and (c) are the measured transmission (OD = 8) in positive and negative scanning $(\Omega_{p} / 2\pi)^{2}$ by setting $T_{s}=2 $ $\mu $s and $T_{s}=18 $ $\mu $s, respectively. (d) and (e) are the measured transmission (OD = 4.5) in positive and negative scanning $(\Omega_{p} / 2\pi)^{2}$ by setting $T_{s}=2 $ $\mu $s and $T_{s}=18 $ $\mu $s, respectively.}
\label{Fig4}
\end{figure*}

The underlying mechanism behind the physical system is based on interaction-induced dissipation, which exhibits how quickly the system respond to changes in probe field. This enables us to capture the time-dependent behavior of system as it undergoes changes in external conditions. In the experiment, we record hysteresis loops versus the scanning time $T_s$ and measure the area enclosed by hysteresis loop. The results are found in Fig.\ref{Fig4}(a). In this scenario, we consider two cases of OD = 8.0 (blue data in Fig.\ref{Fig4}(a)) and OD = 4.5 (red data in Fig.\ref{Fig4}(a)) and show the difference between them. We fit the measured areas using dashed lines, where the fit function is $y = ae^{-b(x-d)^{\alpha}}$+$c$ ($a$ = -64.9, $b$ = 0.053, $c$ = 65.3, $d$ = 5, $\alpha$ = 1.5) for OD = 8.0 and the fit function is $y = 0$ for OD = 4.5. For OD = 4.5, the atoms are dilute, the linearity of transmission on $(\Omega_{p} / 2\pi)^{2}$ is invariant to the scanning time $T_s$ as the interaction is ignored. However, when we increase OD to 8.0, hysteresis loops appear and the area grows versus $T_s$.

In our experiment, a fast scan accumulates a small number of Rydberg atoms within limited time interval to each data, which makes the interactions so weak that the effect from non-Hermiticity might not dominate, see the results shown in Fig.\ref{Fig4}(b). As $T_s$ increases, this corresponds to more excited Rydberg atoms for a relatively large time interval and interactions between the Rydberg atoms cannot be ignored, thus the effect of non-Hermiticity emerges, as illustrated in Fig.\ref{Fig4}(c). For our experiment, there's a characteristic measurement time of $T_s\sim 5 \mu$s, where shorter scan times capture system's transient behavior, while longer scan times converge to the equilibrium value. The results in Figs.\ref{Fig4}(d) and (e) show examples of small OD, the variance of scanning time do not affect the linearity of transmission to probe intensity.

\subsection*{Discussions}
Our experiment serves as a preliminary verification test for non-Hermitian many-body physics \cite{bergholtz2021exceptional,miri2019exceptional}, and promotes the applications towards studying high-order EPs and symmetry breaking in high-dimension systems. For example, according to Ref. \cite{delplace2021symmetry}, both sides of third-order EPs have different signs of winding number, thus providing a platform to study the topological properties (such as topology stability, topological phase transtion) around EPs in the Rydberg system. In addition, the phenomenon of hysteresis loops enables us to build an interface between hysteresis dynamics and non-Hermitian physics, this could provide an experimental correspondence to theory \cite{zhang2020dynamic}. 

In summary, we have observed interaction-induced third-order EPs and hysteresis loops in a cold Rydberg atomic gas. The interaction between Rydberg atoms endows system an dissipation channel, leading to non-Hermitian many-body dynamics. In the experiment, we observe interaction-induced hysteresis loops, in which the dynamics of the system is dramatically distinct by comparing with weak-interaction case. In the context of a cold Rydberg atomic gas, the emergence of EPs and the hysteresis loops due to many-body interactions help us to explore the rich dynamics between many-body interaction and the non-Hermitian physics. 

\begin{figure*}
\centering
\includegraphics[width=1\linewidth]{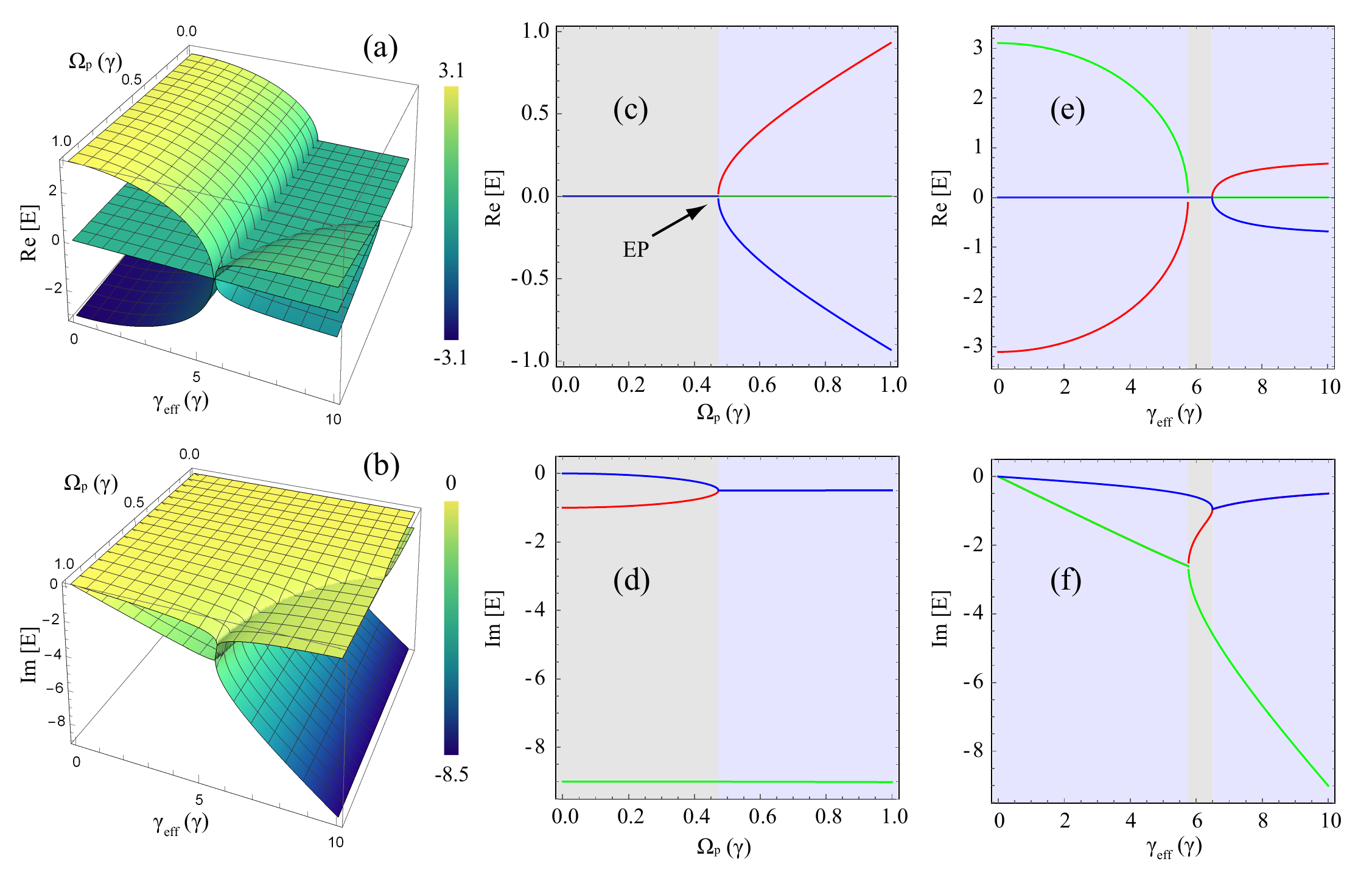}
\caption{\textbf{Theoretical simulations of third-order EPs.} (a) Real and (b) imaginary parts of the system eigenvalues as a function of $\Omega_p$ and $\gamma_\text{eff}$. In these simulations, we set $\Delta_c=\Delta_p=0$, $\Omega_c=3\gamma$. (c) and (d) are the real parts of the system eigenvalues as a function of $\Omega_p$ with $\gamma_\text{eff}=10\gamma$. (e) and (f) represent the system eigenvalues as a function of $\gamma_\text{eff}$ with $\Omega_p=0.8\gamma$. The red, green, and blue lines correspond to the values of $E_1$, $E_2$, and $E_3$, respectively. The gray region in (e) and (f) represents the stable regime of EPs.}
\label{FigS1}
\end{figure*}

\section*{Methods}
\subsection*{Details of the experimental setup}
To study non-Hermitian many-body dynamics, the emergence of EPs, and hysteresis trajectories, we prepare a cold ensemble of $^{85}$Rb atoms trapped in a three-dimensional magneto-optic trap (MOT). The atomic ensemble is prepared in the ground state $\ket{g} = \ket{5S_{1/2}, F = 3, m_{F} = -3}$ by an optical pumping process. In experiment, we wrapped the MOT with a double-layer magnetic shield system. By this way, we can shield the system from external magnetic fields, and it can reduce the internal magnetic field to less than 10 mGauss. A guiding magnetic field is generated by a pair of Helmholtz coils symmetrically placed around the atomic ensemble, and the direction of the field is along the direction of beams propagation. By this way, the direction of the quantisation axis of the system is confirmed. 

We used a two-photon transition scheme to excite $^{85}$Rb atoms from the ground state to the Rydberg state. The probe beam ($\omega_{p} \approx$ 10 µm) driving the atoms from the ground state $\ket{g}$ to the intermediate excited state $\ket{e} = \ket{5P_{3/2}, F = 4, m_{F} = -4}$, and the coupling beam ($\omega_{c} \approx$ 20 µm) then drives the transition from $|e\rangle$ to the Rydberg state $\ket{r} = \ket{47D_{5/2}, F = 5, m_{F} = -5}$, as shown in Fig.\ref{Fig1}(a) in the main text. The probe beam and coupling beam are focused into the cold atomic ensemble, and we used the Pound-Drever-Hall method to lock probe and coupling beams frequency, thus constituting the EIT process. We use a beam splitter to split the coupled beam into two beams before focusing into the cold atomic ensemble, and the intensity of one of these beams was detected using a photodetector. This method allows us to monitor the intensity of the coupling beam in real time. The transmission signals are detected using a photo-multiplier Tube (PMT). 

In experiment, we loaded a triangular wave signal that was generated using a signal generator (RIGOL DG4102) onto the acousto-optic modulator. By this way, we produce the process of increasing beam intensity (Up process) and intensity reduction process (Down process). To better compare the Up and Down processes, we perform an inversion of the Down process, as shown in Fig.\ref{Fig1}(c) in the main text.

\subsection*{Non-Hermitian Hamiltonian}
The system of interest is schematically depicted in Fig.~\ref{Fig1}(b), where two optical fields with spatial overlap comprise a counter-collinear weak-probe field and a strong-control field. The Rydberg atomic level structure is the three-level EIT configuration shown in Fig.~\ref{Fig1}(a). The strong-control field with the Rabi ($\Omega_c$) and detuning ($\omega_c$) field is applied on resonance with the transition $|e\rangle\to
|r\rangle$; while a weak-probe field with the Rabi ($\Omega_p$) and angular ($\omega_p$) frequency is turned almost resonant with the atomic transition $|g\rangle\to
|e\rangle$, but frequency shifted in opposite direction by a small value $\Delta_p$. Under the Rydberg EIT configurations, the population is mainly distributed in $|g\rangle$. The two-photon detuning between two counter-collinear fields is $\Delta=\Delta_p-\Delta_c$, where $\Delta_p$ ($\Delta_c$) is the one-photon detuning of the probe (control) field. The spontaneous decay of the state $|e\rangle$ ($|r\rangle$) at a rate $\Gamma_1$ ($\gamma$). The interactions between excited Rydberg atoms are reflected in the optical responses of atoms and the transmission of the probe field.

The Hamiltonian in the interaction picture and rotating-wave approximation reads $(\hbar=1)$,
\begin{equation}
\begin{aligned}
H_{many} &=
\sum_{j=1}^{N} \left[-(\Delta_p+\Delta_c)\hat{\sigma}_{rr}^{j}-\left(\Omega_p\sigma_{eg}^j+\Omega_c\sigma_{re}^j +{\rm H.c.}\right)\right]\\
&+\sum_{j=1}^{N} \left[-\Delta_p\hat{\sigma}_{ee}^j +\sum_{j<k}V_{jk}\sigma_{rr}^j\sigma_{rr}^k\right],
\end{aligned}
\end{equation}
where $\sigma^{j}_{\alpha\beta}=|\alpha_j\rangle\langle \beta_j|$ ($\alpha, \beta = e, g, r$). We then consider the mean-field approximation, in which a single atom is immersed in a field generated by the interactions between itself and other atoms. Consequently, the problem of solving the dynamic many-body system is reduced to addressing the dynamics of one single atom within that field, treating the other atoms as part of the environment. The Hamiltonian can be written as 
\begin{equation}
    H_{many}=H_{1}\otimes I_{N-1}+I_{1}\otimes H_{N-1}+H_{I},
\end{equation}
where $H_{I}$ describes the interaction between the single atom and the environment, $H_{1}$ is the single atom Hamiltonian, $H_{N-1}$ is the Hamiltonian of the environment, and $I_{1}$ ($I_{N-1}$) denotes the identity in the Hilbert space $H_{1}$ ($H_{N-1}$). Consider the two-body interaction between the Rydberg atoms, the interaction distance $R_j$ represents the single atom between the atom in the environment. Then the total Rydberg state $|r_{N}\rangle=\sum_{j=1}^{N-1}|r_{1}\rangle\otimes|r_{j}\rangle$ evolves according to 

\begin{equation}
    \hat{U}(t)=\sum_{j=1}^{N-1} e^{-iV_{j}t} \left( |r_{1}\rangle\otimes \sum_{i=1}^{N-1}|r_{i}\rangle_{j} \right)\left( \langle r_{1}|\otimes \sum_{i=1}^{N-1}\langle r_{i}|_{j} \right).
\end{equation}

Since the distance between the single atom and the $j$-atom in the environment $R_j$ is different, in regarding the energy shifts on $|r_{1}\rangle$, it is necessary to consider the distinct contributions from the $j$-atom. The atom far away in the environment produces a small shift, otherwise the shift will be large, due to the form of van der Waals (vdW) potential $V_{\text{vdW}}\propto{C_6}/{R^6}$. Overall, this effect is to widen the energy level $|r_{1}\rangle$ with an effective width $ \gamma_{\rm eff}$. Here, we consider the reduced density matrix of the single-atom system, and the non-Hermitian Hamiltonian has the form
\begin{equation}
\begin{aligned}
H_{1} &=-\Delta_p\hat{\sigma}_{ee}
-(\Delta_p+\Delta_c-\frac{i \gamma_{\rm eff}}{2})\hat{\sigma}_{rr}\\
&-\left(\Omega_p\sigma_{eg}+\Omega_c\sigma_{re}+{\rm H.c.}\right)
\end{aligned}
\end{equation}
where $\gamma_{\rm eff}$ reveals the effective decay rate induced by the interaction of the environment. The eigenvalues of $H_{1}$ are given by 
\begin{equation}
E_{1}=\frac{1}{6} \left(\frac{C}{\sqrt[3]{B+i A}}-\sqrt[3]{4B+4i A}-2 i \gamma _{\text{eff}}\right)
\end{equation}
\begin{equation}
E_{2}=\frac{1}{12} \left(\left(1-i \sqrt{3}\right) \sqrt[3]{4B+4i A}-\frac{\left(1+i \sqrt{3}\right) C}{\sqrt[3]{B+i A}}-4 i \gamma _{\text{eff}}\right)
\end{equation}
\begin{equation}
E_{3}=\frac{1}{12} \left( \left(1+i \sqrt{3}\right) \sqrt[3]{4B+4i A}-\frac{\left(1-i \sqrt{3}\right) C}{\sqrt[3]{B+i A}}-4 i \gamma _{\text{eff}}\right)
\end{equation}
where
\begin{equation}
A=9 \gamma _{\text{eff}} \left(\text{$\Omega $c}^2-2 \text{$\Omega $p}^2\right)-2 \gamma _{\text{eff}}^3
\end{equation}
\begin{equation}
B=3 \sqrt{3} \sqrt[3]{ \begin{split}&\gamma _{\text{eff}}^2 \left(\Omega_c^4+20\Omega_c^2 \Omega_p^2-8\Omega_p^4\right)\\
&-4\Omega_p^2 \gamma _{\text{eff}}^4-4 \left(\Omega_c^2+\Omega_p^2\right)^3\end{split}}
\end{equation}
\begin{equation}
C=2 \sqrt[3]{2} \left(\gamma _{\text{eff}}^2-3 \left(\text{$\Omega $c}^2+\text{$\Omega $p}^2\right)\right)
\end{equation}
By setting the parameters $\Omega_c=3\gamma$ and $\Delta _p=\Delta_c=0$, we obtain the real and imaginary parts of the system eigenvalues as a function of $\gamma_\text{{eff}}$ and $\Omega_p$, as given in Fig.\ref{FigS1}. From these results, for example in Fig.~\ref{FigS1}(c), we can find that the third-order EPs appear where the real parts of three eigenvalues coalesce simultaneously. At the cases in Figs.~\ref{FigS1}(c) and (d), the real and imaginary parts of the eigenvalue $E_2$ are constant. In addition, when $\Omega_p=0.8\gamma$, the third-order EPs first appear, and retain stable in a range of $5.77\gamma<\gamma_\text{eff}<6.49\gamma$, and then disappear. This suggests that there are regions in the parameter space where certain conditions are met to create this high-order degeneracy. 

\begin{figure*}
\centering
\includegraphics[width=1\linewidth]{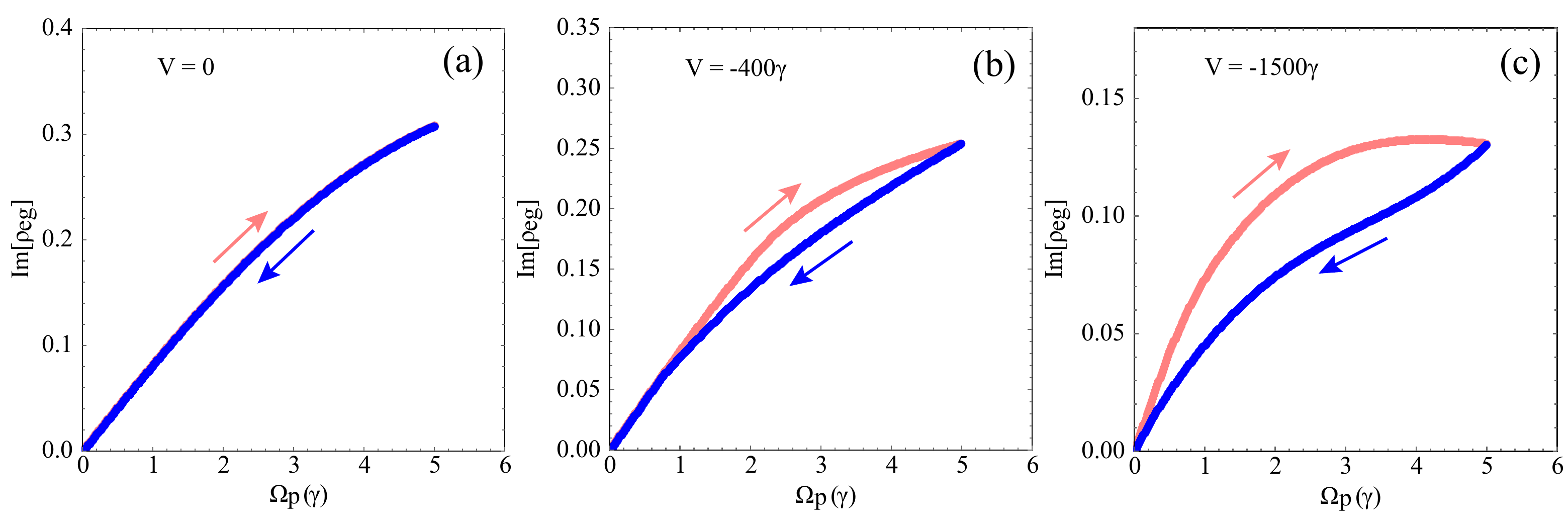}
\caption{\textbf{Hysteresis trajectories of Im[$\bf{\rho_{eg}}$] versus interaction strength $V$.} The simulated hysteresis versus interaction strength are given in (a) V = 0, (b) V = -400 $\gamma$, and (c) V = -1500 $\gamma$. The pink and blue arrows represents the positive and negative scanning $\Omega_p$. In these simulations, we set $\Gamma_1 = 12\gamma$, $\Delta_c=\Delta_p=0$, $\Omega_c = 0.25\gamma$, and $\rho _{re}(t=0)$=0.1.}
\label{FigS2}
\end{figure*}

\subsection*{Lindblad master equation}
In the mean-field treatment, we solve the Lindblad master equation by adding the non-Hermitian term $\Gamma_2\rightarrow \gamma+\gamma_{\rm eff}$, and obtain the following equations:
\begin{align}
{\dot \rho _{gg}}&=i \frac{\Omega_p}{2} (\rho_{eg} - \rho_{ge}) + \Gamma_1 \rho_{ee},\\
{\dot \rho _{ee}}&=i \frac{\Omega_p}{2} (\rho_{ge} - \rho_{eg}) + i \frac{\Omega_c}{2} (\rho_{re} - \rho_{er})- \Gamma_1 \rho_{ee} + \Gamma_2 \rho_{rr}, \\
{\dot \rho _{rr}}&=i \frac{\Omega_c}{2} (\rho_{er} - \rho_{re})- \Gamma_2 \rho_{rr},\\
{\dot \rho _{re}}&=i \frac{\Omega_c}{2} (\rho_{ee} - \rho_{rr}) - i \frac{\Omega_p}{2} \rho_{rg}- \left(i \Delta_c + \frac{\Gamma_2}{2}+\frac{\Gamma_1}{2}\right) \rho_{re}, \\
{\dot \rho _{er}}&=i \frac{\Omega_c}{2} (\rho_{rr} - \rho_{ee}) + i \frac{\Omega_p}{2} \rho_{gr} + \left(i \Delta_c - \frac{\Gamma_2}{2} - \frac{\Gamma_1}{2}\right) \rho_{er},\\ 
{\dot \rho _{rg}}&=i \frac{\Omega_c}{2} \rho_{eg} - i \frac{\Omega_p}{2} \rho_{re} - \left(i (\Delta_p + \Delta_c) + \frac{\Gamma_2}{2}\right) \rho_{rg}, \\
{\dot \rho _{gr}}&= i \frac{\Omega_p}{2} \rho_{er}-i \frac{\Omega_c}{2} \rho_{ge} +\left(i (\Delta_p -\Delta_c)-\frac{\Gamma_2}{2}\right) \rho_{gr},\\
{\dot \rho _{eg}}&=i \frac{\Omega_p}{2} (\rho_{gg} - \rho_{ee}) + i \frac{\Omega_c}{2} \rho_{rg} - \left(i \Delta_p + \frac{\Gamma_1}{2}\right) \rho_{eg}, \\
{\dot \rho _{ge}}&=i \frac{\Omega_p}{2} (\rho_{ee} - \rho_{gg}) - i \frac{\Omega_c}{2} \rho_{gr} + \left(i \Delta_p - \frac{\Gamma_1}{2}\right) \rho_{ge}, 
\end{align}

First, we consider the case of no-interaction ($\gamma_{\rm eff}=0$) and calculate the steady-state solution ($\dot \rho _{ij}$=0, where $i,j$ represent the states of  $\ket{g}$, $\ket{e}$, and $\ket{r}$), we therefore obtained the element of density matrices $\rho_{eg}$ and $\rho_{rr}$ versus $\Delta_p$ and $\Gamma_2$, with forms of $\rho_{eg}$($\Delta_p$, $\gamma$) and $\rho_{rr}$($\Delta_p$, $\gamma$), respectively. Then, by considering the interaction induced non-Hermiticity, we obtain the modified matrix element $\rho_{eg}$($\Delta_p$, $\gamma$ + $V\rho_{rr}$) with the mean-field approximation $\Gamma_2 \rightarrow \gamma + V\rho_{rr}$. We plot Im[$\rho_{eg}$] versus $\Delta_p$ and $V$, the phase diagram and transmission lines are given by Fig.\ref{Fig2.2}. 

To simulate hysteresis trajectories, the state of atoms is influenced not only by the external input but also by their previous states. In particular, the systematic evolution is dependent on the direction of scanning $\Omega_p$. Thus, we replace the time-dependent population $\rho_{rr}(t)$ by the term $\dot \rho _{rr}(t)$ according to Eq. (14): $ \rho_{rr}(t)=i \frac{\Omega_c}{2\Gamma_2} (\rho_{er}(t) - \rho_{re}(t))-\frac{\dot \rho _{rr}(t)}{\Gamma_2}$. Thus, the slope of Rydberg population $\dot \rho _{rr}(t)$ plays a crucial role in influencing the transient behavior of $\rho_{rr}(t)$. 

In the simulations, we treat the results $\rho_{ij}(t)$ at time $t$ as the initial conditions for calculating $\rho_{ij}(t+\Delta t)$ at time $t+\Delta t$ . This approach effectively captures the inherent memory effects in the system's evolution. We specifically examined the time-dependent component Im[$\rho_{\rm{eg}}(t)$] while scanning $\Omega_p$ both positively and negatively, across various interaction strengths: $V=0$, $V=-400 \gamma$, and $V=-1500 \gamma$. The corresponding results are presented in Figs.\ref{FigS2} (a), (b) and (c). These results reveal intriguing behaviors of the scanned trajectories. In the case with no interaction (\(V = 0\), Fig. \ref{FigS2} (a)), the trajectories exhibit a coincident pattern, indicating that the system's dynamics are symmetric. However, when considering interactions, specifically at $V = -400 \gamma$ and $V = -1500 \gamma$, as depicted in Figs. \ref{FigS2} (b) and (c), the trajectories form closed loops.

The emergence of these hysteresis loops underscores the critical role of interaction strength in shaping the dynamical properties of the system. Moreover, the observed dynamics align closely with experimental observations illustrated in Figs. \ref{Fig3} (a-d) and Figs. \ref{Fig4} (a-e), which further corroborates the relevance of our simulation results. This consistency between theoretical predictions and experimental observations highlights the exact complex interplay between system memory and interaction, providing valuable insights into the underlying mechanisms governing the observed phenomena.

\section*{Acknowledgements}
We acknowledge funding from the National Key R and D Program of China (Grant No. 2022YFA1404002), the National Natural Science Foundation of China (Grant Nos. U20A20218, 61525504, and 61435011), the Anhui Initiative in Quantum Information Technologies (Grant No. AHY020200), and the Major Science and Technology Projects in Anhui Province (Grant No. 202203a13010001).

\section*{Author contributions statement}
D.-S.D. conceived the idea with discussion from E.Z.L. J.Z and Y.J.W conducted the physical experiments. D.-S.D. and E.Z.L developed the theoretical model. The manuscript was written by D.-S.D, J.Z., E.Z.L and Y.J.W.  The research was supervised by D.-S.D. All authors contributed to discussions regarding the results and the analysis contained in the manuscript.

\section*{Competing interests}
The authors declare no competing interests.

\bibliography{ref}

\end{document}